# The cost of clean air: a price analysis of air filtration technology

Andrew Yang, Mary Prunicki, Kari Nadeau

The Geffen Academy at UCLA

Stanford Sean N. Parker Center of Allergy and Immunology Research

## Abstract

Air pollution is a pernicious but increasingly rampant health problem in human populations across the globe. Varying duration of exposure to airborne pollutants can result in a plethora of health conditions, ranging from short-term bronchitis to prolonged cardiovascular diseases and lung cancers. Under chronic exposure to pollution, medical treatment for pollutant-caused health conditions can be quite costly. The most immediate and effective way to combat the health detriments of pollution is air purifiers, which take in ambient air and filter out the harmful pollutants before releasing it back into the surroundings. Filters are highly effective at decreasing air pollution, and by association, hospitalizations from exposure. An analysis of 52 common air purifiers was undertaken to compare acquisition and maintenance costs during severe air pollution levels across four counties in California. This approach enables the "cost" of clean air to be put into perspective i.e., the price compared to the cost of healthcare and other expenses, population age, and pre-existing conditions.

## 1. Introduction

In recent years, air pollution has become a growing problem [1]. With the advent of industrialization and increased urbanization, air pollution is a health hazard to those who live in and around cities. In addition to human pollution, smoke and dust from natural events such as wildfires and dust storms worsen air quality, affecting wide areas [1][2].

In California, pollution from urban environments compounded by frequent wildfires leads to poor air quality affecting over 38 million residents in 2021 [3]. For populations more at risk, such as the elderly, children, and those with preexisting lung conditions, poor air quality is especially of concern [1][2]. The health effects of air pollution generate costly medical expenses [3].

However, air purifiers remove air pollutants and therefore reduce air pollution caused medical costs [4][5]. The goal of this paper is to determine how the cost of investing in an air purifier compares to medical expenses of sensitive populations. In Section 2, a review of pollutants and filtering technology is provided. In Section 3, a cost model is developed to compare different air purifier models in an objective fashion by taking price and efficacy into account. In Section 4, the cost model is then applied to evaluate air purifiers running in environments with varying levels of air quality. The cost model results are analyzed for trends, then compared to health expenses due to air pollution based on data gathered from previous studies.

## 2. Background

The term "air pollution" used in literature and the popular press is broadly inclusive of many chemicals and particulate matter. Such pollutants have various health effects, and air purifiers vary in their ability to filter these different particles. This section first reviews different sources and compositions, followed by a brief description of health effects, and then describes the operation and capability of air purifiers.

### 2.1 Makeup of Air Pollution

Air pollution is comprised of various chemicals and particles from numerous sources. The pollutants are generally classified into three categories, gaseous pollutants, particulate matter, and microorganisms [6][7].

Simple gaseous pollutants include carbon dioxide ($CO_2$), sulfur dioxide ($SO_2$), CO (carbon monoxide), and $NO_x$ (nitrogen oxides). They originate from power generators and agriculture, as well as transportation exhaust. Another class, volatile organic compounds (VOC), include carcinogens such as benzene and formaldehyde, that result from fuel combustion, industrial processes, and pesticides [1][7]. Heavy metals are rarer in the atmosphere but include toxic metals such as mercury, lead, and cadmium that result from industrial processes. These tend to build up in patients slowly and are difficult or unable to be excreted from the body [1][7].

The second broad classification is particulate matter (PM) which are large particles, either solid or liquid, floating in the air that may or may not be reactive with the human body. It is classified by size in micrometers. For instance, PM 2.5 indicates fine particulate matter in the air that has a diameter of less than 2.5 µm. The finer the particulate matter, the easier it is to be absorbed into the bloodstream after inhalation. PM often contains reactive chemicals, including the types listed above [6].

Microorganisms are the third category that include viruses, spores, and bacteria. They are typically produced by living organisms and are often embedded in larger particles or droplets during airborne transmission [7]. Air purification methods like HEPA can not remove the microorganisms directly due to their small size but can remove the larger particles that microorganisms are suspended in.

A summary of the size of air pollutants is shown in Table 1.

**Table 1**

*Sizes of common air pollutants*

| Particle | Particle Size (microns) |
|---|---|
| Atmospheric Dust | 0.001 – 40 |
| Bacteria | 0.3 – 60 |
| Beach Sand | 100 – 10000 |
| Burning Wood | 0.2 – 3 |
| Cement Dust | 3 – 100 |
| Clay, fine | 0.5 – 1 |
| Coal Dust | 1 – 100 |
| Combustion | 0.01 – 0.1 |
| Dust Mites | 100 – 300 |
| Fly Ash | 1 – 1000 |
| Grain Dusts | 5 – 1000 |
| Household dust | 0.05 – 100 |
| Human Hair | 40 – 300 |
| Insecticide Dusts | 0.5 – 10 |
| Lead Dust | 0.1 – 0.7 |
| Mold Spores | 10 – 30 |
| Pet Dander | 0.5 – 100 |
| Pollen | 10 – 1000 |
| Smoke | 0.01 – 0.1 |
| Tobacco Smoke | 0.01 – 4 |
| Viruses | 0.005 – 0.3 |

*Note*. Adapted from Breathe Quality, "What is True HEPA filter? True HEPA filter vs HEPA-type filter", 2019.

## 2.2     Health Effects of Air Pollution

The health effects of human exposure to air pollution are well documented to be detrimental. Short-term exposure, classified as duration of fewer than 30 days, is linked to COPD (chronic obstructive pulmonary disease) and exacerbates preexisting conditions such as asthma and emphysema. Even on days with lower air pollution, exposure can induce coughing and wheezing, and increased rates of hospitalization [1][2][10].

Effects of long-term exposure are dependent on a myriad of factors, including the chemical composition of gas or particles, or the vitality of the exposed patient. Gaseous pollutants often lead to an increased risk of heart attacks from various cardiovascular diseases and permanent accumulation of lung lesions and decreased pulmonary function [2][1][8]. Maternal exposure to airborne toxins is correlated to reduced fetal growth and impaired cognitive development, as well as other detrimental respiratory, cardiovascular, and perinatal disorders in early human life. [8]. Particulate matter has a bigger influence on people with preexisting conditions such as asthma or eczema.  In all cases, any exposure to air pollution, independent of type of pollutant, duration, and preexisting health conditions, lead to increased rates of mortality and hospitalization [1].

## 2.3    Air Purifier Technology

Air purifiers are designed to have a combination of technologies to remove different pollutants so that in combination clean air delivery is ensured. A purifier uses an electric motor driving a fan that forces the ambient air through one or more air filters of differing compositions. The air in any given environment is continually recycled through the air purifier. The ability of the purifier to clean air depends on the volume of air through the filters and the ability of the filters to eliminate the pollutant [9].

The design of the motor and fan in an air purifier depends on the target airflow measured in cubic feet per minute (CFM). Industrial environments and clean rooms have large centralized systems that drive an extremely high volume of air through the building filtration system to recycle the room's air up to 30 times per hour (air change per hour) [10]. Home-use air purifiers are stand-alone units and hence integrate their own fans. Depending on the size of the unit, typical home air purifiers provide airflow of several hundred CFM with fans needing less than 100W. Higher wattage would require higher operating electrical costs and may be incompatible with home electrical systems.

The basis of air filtration is an air filter or cartridge made with tightly woven fibers that catch fine particles when ambient air passes through [9]. In order for these air purifiers to remain effective, these cartridges need to be replaced after filtering a certain amount of air. High efficacy particulate air (HEPA) filters are the most common example of this filtration technology available for home environments. The difference between these filters is the ratings for removing PM. The table below shows two common numerical ratings, HEPA and minimum efficiency rating values (MERV), and the associated efficiency in removing PM of varying dimensions. Filters with HEPA ratings of 15 and above are considered ultra-low particulate air (ULPA) filters targeting clean room or industrial environments. In general, HEPA rates of 12 and below and MERV ratings of 16 and below are typically used for home environments. While air filters with such ratings are available, it is well recognized that HEPA filters guarantee the highest decrease of indoor PM, especially taking affordability and consistency into account [4][11][9].

**Table 2**

*Percent efficacy of HEPA and ULPA ratings*

| HEPA class | Efficiency |
| --- | --- |
| E10/H10 | 85% |
| E11/H11 | 95% |
| E12/H12 | 99.5% |
| H13 | 99.97% |
| H14 | 99.975% |
| U15 | 99.9975% |
| U16 | 99.99975% |
| U17 | 99.9999% |

*Note*. Adapted from Breathe Quality, "What is True HEPA filter? True HEPA filter vs HEPA-type filter", 2019.

**Table 3**

*MERV ratings in terms of dust filtering efficiency and minimum particle size*

| MERV rating | Dust Efficiency | Particle size |
| --- | --- | --- |
| 20 | ≥ 99.999% | 0.1 – 0.2 microns |
| 19 | ≥ 99.99% | 0.1 – 0.2 microns |
| 18 | ≥ 99.97% | 0.1 – 0.2 microns |
| 17 | ≥ 99.97% | 0.3 microns |
| 16 | ≥ 99.95% | 0.3 – 1 micron |
| 15 | ≥ 95% | 0.3 – 1 micron |
| 14 | 90 – 95% | 0.3 – 1 micron |
| 13 | 89 – 90% | 0.3 – 1 micron |
| 12 | 70 – 75% | 1 – 3 microns |
| 11 | 60 – 65% | 1 – 3 microns |
| 10 | 50 – 55% | 1 – 3 microns |
| 9  | 40 – 45% | 1 – 3 microns |
| 8  | 30 – 35% | 3 – 10 microns |
| 7  | 25 – 30% | 3 – 10 microns |
| 6  | < 20%    | 3 – 10 microns |
| 5  | < 20%    | 3 – 10 microns |
| 4  | < 20%    | ≥ 10 microns |
| 3  | < 20%    | ≥ 10 microns |
| 2  | < 20%    | ≥ 10 microns |
| 1  | < 20%    | ≥ 10 microns |

*Note*. Adapted from Breathe Quality, "What is True HEPA filter? True HEPA filter vs HEPA-type filter", 2019.

Another class of filters, activated charcoal filters, are able to remove reactive chemicals. It is most commonly employed to purify the air of gaseous pollutants including VOCs .Due to the high surface area created by activated charcoal, it acts like a sponge to adsorb chemicals. While activated charcoal filters are able to adsorb pollutants, quantifying the effect seems to be inconsistently measured across studies [9]. Because of the low cost of adding such a filter to the overall air purifier, they are typically included.

An alternative to using woven fibers is to use an electrostatic filter in the air purifier. The purifiers create a static charge along a filter surface that attracts pollutants, thereby removing them from the air. A key advantage to these filters is that no air filter cartridges are needed and wiping the filter surface periodically restores the efficacy. In ideal environments, they have been shown to be able to remove a very wide range of different-sized particles and chemical pollution [9]. However, Liu et al. have shown that electrostatic filtration cleans air very slowly, and often achieves a 50% or less efficacy rate.

Finally, ultraviolet (UV) filters are additionally employed against microbes and living organisms. Instead of catching particles, the air is bathed in UV light. The high-energy light breaks down the genetic material of the microbes, killing the microorganism. This technology is not applicable to any nonbiological pollutants [9]. One risk of using electrostatic and UV air filters is ozone production from reacting with ambient air. Ozone irritates the lungs and is considered an air pollutant. The EPA (environmental protection agency) warns specifically against using such devices.

An effective filtration pairing commonly found in products is a HEPA filter with activated carbon [9]. The HEPA filter catches PM> 0.3um, while gaseous pollutants that slip through the HEPA fibers are caught by the activated charcoal filter. Because the other filtration technologies, i.e. electrostatic and UV, have been shown to be potentially harmful or ineffective, this study focuses on air purification utilizing this combination. This level of filtration is clearly sufficient for removing PM 2.5 particles that are of common concern and measured by agencies such as EPA.

## 3. Cost Model of Air Purifiers

Numerous standalone air purifiers for home use are available from many manufacturers, each with multiple models. The choices can be overwhelming for individuals or families seeking a clean home air environment. The choice is exacerbated by the varying claims of manufacturers in terms of affordability and efficiency. This section selects a list of credible manufacturers and purifiers models and normalizes their performance and introduces a cost model to enable comparison.

### 3.1  Purification Cost per Year

This model accounts for several factors. While the initial price is important, the actual cost of ownership also requires operating costs which include both the replacement filters to be installed after a certain duration of usage and the cost of electrical power to operate. The paper introduces a figure of merit of Purification Cost per Year (PCY):

$$PCY (\$/year) = (C_{initial}/10 + C_{maintenance/yr} + C_{elec/yr})*(2500/A_{optimal}).$$

$C_{initial}$ refers to the initial cost of the unit. The calculation assumes that the unit needs to be replaced after 10 years. The $C_{maintenance/yr}$ can be expressed as follows and refers to the cost of replacing filters in one year.

$$C_{maintenance/yr} = (T_{operate}*C_{filter}/T_{replacement})$$

$C_{filter}$ is the cost of each filter. $T_{replacement}$ is the suggested duration (in days) by the manufacturer to keep the unit filtrating properly. $T_{operate}$ is the number of days the unit is operating per year. $C_{elec/yr}$ is the cost of electricity from operating the air purifier in a year also based on $T_{operate}$. And, $A_{optimal}$ is the optimal area covered by the air purifier. The metric is normalized to an average home size of 2500 square feet.

This metric depends on location because $C_{elec/yr}$ needs to account for the local cost of electricity, $C_{local}$. $C_{elec/yr}$ can be expressed as

$$C_{elec/yr} = T_{operate} * (W/1000) * C_{local} * 24$$

$C_{local}$ varies widely even within the United States. Table 4 below shows a sampling of states and their average electrical cost. The overall average within the US is 14 cents/kWh and can range from 10.1 cents/kWh to 30.4 cents/kWh.

**Table 4**

*Electricity costs across a selection of representative states*

| State | ¢/kWh |
|---|---|
| New York | 14.3 |
| Texas | 12.8 |
| Alaska | 22.7 |
| Michigan | 16.3 |
| Washington | 10.0 |
| Florida | 12.0 |
| New Mexico | 13.4 |
| Colorado | 12.3 |
| Hawaii (highest) | 30.4 |
| Louisiana (lowest) | 10.1 |
| Average in United States | 14.5 |

*Note*. Adapted from U.S Energy Information Administration, "State Electricity Profiles". https://www.eia.gov/electricity/state/

Our results table will calculate this cost using the rate of California (25.1 cents/kWh), a state with poor air due to regular wildfires.

The PCY metric includes the area covered by an air purifier, $A_{optimal}$. This value varies greatly between manufacturers and models. The value for a metric is calculated based on clean air delivery rate (CADR). CADR s an external rating of air purifier efficacy as established by the Association of Home Appliance Manufacturers(AHAM). Only models with CADR are included in the final results table. CADR is calculated by letting an air purifier run in a room filled with a predetermined concentration of pollutants, then measuring the concentration before and after filtration. The AHAM recommends using air purifiers that have a CADR of ⅔ the size of the room being filtered.

Therefore, in accordance with AHAM guidelines, the advertised coverage of an air purifier is not used in the PCY to avoid inaccurate manufacturer claims. In its place, the optimal coverage is calculated as follows:

$$A_{optimal} = CADR * 3/2$$

Table 5 shows results for 54 air purifier units. These units are selected based on the completeness and availability of information. The initial cost of purchasing the unit and the cost of the replacement filters are based on the price on Amazon. Since Amazon is the largest retailer within the US, the pricing can be considered fair for each unit. These 54 units are from 12 companies: Coway, IQAir, Bissell, Winix, Alen, Hathaspace, Medify, Veva, Air Doctor, PureZone, Honeywell, Blueair. These companies represent a diverse price range and coverage area. Some models within each brand represent older model years and are included in the analysis to evaluate the impact of improved technology on the PCY metric.

**Table 5**

*PCY of 52 air purifier units*

| Product | $C_{initial}$ | Lifetime $C_{filter}$ | Lifetime $C_{elec}$ | $A_{optimal}$ | PCY |
|---|---|---|---|---|---|
| Coway Airmega 250 | $290.88 | $378.54 | $1,341.24 | 372 | $1,155.77 |
| Coway Airmega 150 | $146.35 | $499.56 | $769.57 | 207 | $1,532.76 |
| Coway Airmega AP-1512HH | $229.99 | $378.54 | $1,693.05 | 369 | $1,403.51 |
| Coway Airmega 400 | $400 | $1,289.12 | $1,407.21 | 525 | $1,283.96 |
| Coway Airmega 300 | $498 | $643.56 | $1,231.31 | 510 | $919.05 |
| Coway Airmega AP-1216L | $178.58 | $574.51 | $1,083.99 | 352.5 | $1,176.24 |
| IQAir HealthPro Compact | $799 | $1,411.82 | $2,968.33 | 495 | $2,212.19 |
| IQAir GC MultiGas | $1299 | $914.97 | $8,795.04 | 450 | $5,394.45 |
| IQAir HealthPro Plus | $899 | $1,411.82 | $2,968.33 | 450 | $2,433.41 |
| IQAir Atem Desk | $399 | $689.53 | $74.76 | 27 | $7,076.72 |
| IQAir Atem Car | $399 | $789.46 | $74.76 | 19.5 | $11,079.71 |
| Bissell air220 | $236.89 | $379.54 | $1,649.07 | 265.5 | $1,910.18 |
| Bissell air320 | $339.99 | $379.54 | $2,550.56 | 367.5 | $1,993.27 |
| Bissell air400 | $360.49 | $519.44 | $1,649.07 | 418.5 | $1,295.41 |
| Winix 5500-2 | $249.99 | $524.14 | $1,978.88 | 369 | $1,695.82 |
| Winix AM90 | $219.99 | $406.02 | $1,429.19 | 360 | $1,274.46 |
| Winix 5300-2 | $199.99 | $406.02 | $1,539.13 | 369 | $1,317.85 |
| Winix HR900 | $215.99 | $406.02 | $1,978.88 | 451.5 | $1,320.55 |
| Winix A231 Tower | $99.99 | $499.06 | $1,209.32 | 220.5 | $1,936.93 |
| Winix XLC | $499.95 | $1,488.98 | $1,978.88 | 600 | $1,444.94 |
| Winix C545 | $249 | $406.02 | $1,429.19 | 369 | $1,243.37 |
| Alen 75i | $759 | $689.53 | $989.44 | 520.5 | $806.42 |
| Alen BreatheSmart Classic | $649 | $789.46 | $2,308.70 | 450 | $1,721.20 |
| Alen 45i | $428.99 | $689.53 | $1,099.38 | 367.5 | $1,216.94 |
| Alen FLEX | $349 | $689.53 | $989.44 | 337.5 | $1,243.68 |
| Alen T500 | $213.99 | $639.56 | $1,165.34 | 213 | $2,118.43 |
| Alen FIT50 | $579 | $889.39 | $1,319.26 | 330 | $1,673.22 |
| Hathaspace HSP001 | $299 | $1,198.88 | $879.50 | 240 | $2,164.98 |
| Hathaspace HSP002 | $399.99 | $2,248.16 | $2,198.76 | 675 | $1,647.01 |
| Medify MA-40 | $349 | $719.51 | $1,495.16 | 495 | $1,118.52 |
| Medify MA-25 | $160 | $749.49 | $615.65 | 202.5 | $1,685.36 |
| Medify MA-14 | $97.5 | $599.59 | $571.68 | 180 | $1,626.76 |
| Medify MA-50 | $499 | $1,349.08 | $1,209.32 | 450 | $1,421.33 |
| Medify MA-112 | $594.99 | $1,678.85 | $2,088.82 | 1425 | $661.00 |
| Medify MA-35 | $274.99 | $2,098.26 | $879.50 | 255 | $2,919.38 |
| Medify MA-18 | $95 | $749.49 | $769.57 | 225 | $1,687.84 |
| Medify MA-CAR | $98.99 | $374.74 | $105.54 | 30 | $4,002.37 |
| Medify MA-22 | $119 | $599.59 | $835.53 | 225 | $1,594.58 |
| VEVA 8000 Elite Pro | $94.99 | $139.80 | $1,099.38 | 202.5 | $1,529.86 |
| AirDoctor 3000 | $495 | $1,448.51 | $2,418.64 | 514.5 | $1,879.08 |
| AirDoctor 1000 | $399 | $1,598.70 | $879.50 | 228 | $2,717.33 |
| AirDoctor 5000 | $999 | $3,547.57 | $2,198.76 | 801 | $1,793.49 |
| PureZone Elite | $149.99 | $299.69 | $1,319.26 | 180 | $2,248.54 |
| PureZone Mini | $39.99 | $199.66 | $164.91 | 7.5 | $12,152.34 |
| PureZone Breeze | $79.99 | $399.53 | $725.59 | 67.5 | $4,167.10 |
| PureZone Halo | $99.99 | $299.59 | $879.50 | 75 | $3,930.33 |
| Honeywell HPA300 | $249.99 | $249.73 | $2,858.39 | 450 | $1,726.73 |
| Honeywell HPA200 | $199.99 | $998.82 | $1,846.96 | 300 | $2,371.48 |
| Honeywell HPA100 | $134.99 | $109.82 | $1,099.38 | 150 | $2,015.34 |
| Blueair Pure 211+ Auto | $339.99 | $1,398.84 | $835.53 | 570 | $979.99 |
| Blueair Pure 211+ | $319.99 | $839.23 | $1,341.24 | 525 | $1,038.32 |
| Blueair Pure 311 Auto | $249.99 | $414.72 | $769.57 | 375 | $789.52 |
| Blueair Pro XL | $2499.99 | $1,398.84 | $5,628.83 | 1200 | $1,464.10 |

*Note.* $A_{optimal}$ is listed in $ft^2$. "Lifetime" implies that $C_{elec}$ and $C_{filter}$ were calculated across a span of 10 years, signifying the lifetime of an air purifier unit. All calculations assume constant air purifier usage.

## 4. Results and Discussion

Several observations can be made based on the results in Table 5 of the previous section. This section evaluates the potential cost of clean air and the comparison to medical cost as a reference.

### 4.1 Cost of Clean Air

The results of Table 5 assume continuous operation for an entire year in California, out of the 54 air purifiers assessed in Table 5, the median PCY is $1607.52 per year. Under the same conditions while varying the state, the range across the United States was $1002.12 based on electricity cost.

The PCY can be broken into its three component cost contributors: $C_{initial}/10$, $C_{elec/yr}$, and $C_{maintenance/yr}$. The cost breakdown of PCY into its respective factors is depicted in Figure 1.

**Figure 1**

*Percentage breakdown of PCY into cost components*

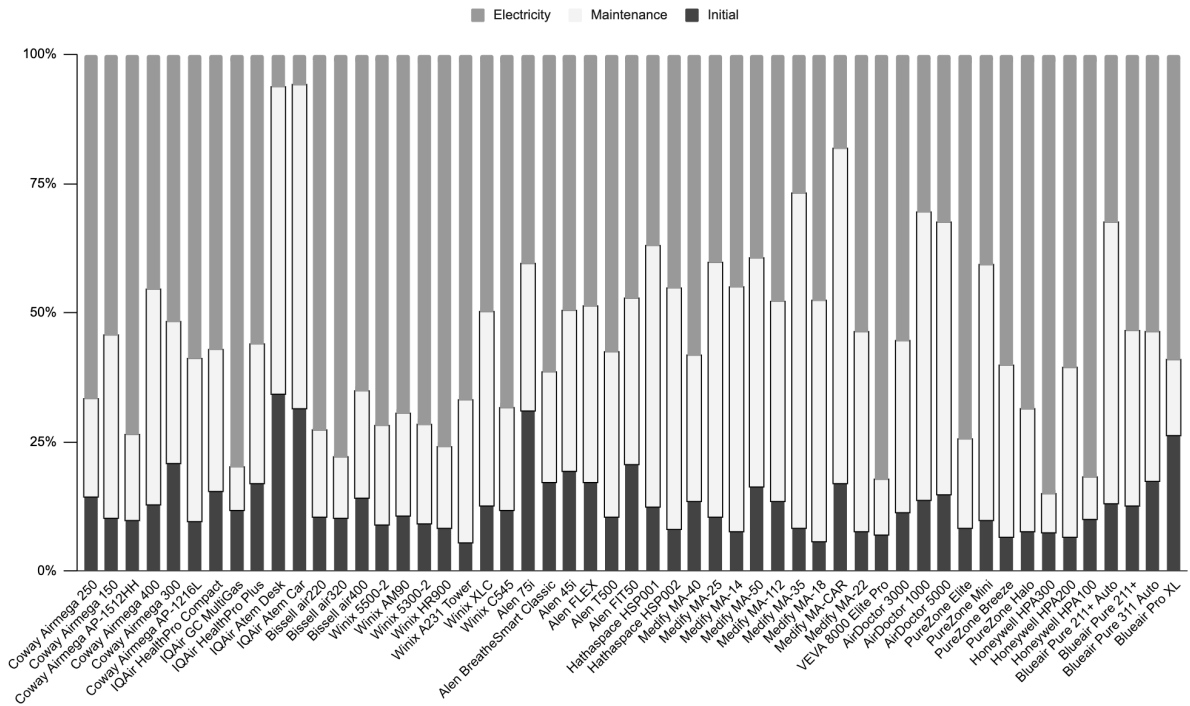

Initial cost as a percentage of PCY remains relatively constant. On average, the buying cost of a purifier spread over 10 years makes up 13.8% of the total expenses generated in one year of use. Alternatively put, the initial cost of a purifier is averaged around 1/7 of the total costs generated in a 10-year purifier lifetime.

The percentage of the other cost factors ($C_{elec/yr}$, and $C_{maintenance/yr}$) were uniformly distributed but yielded no statistical significance.

### 4.2   PCY Versus Cost Components

Coverage varied inversely with lifetime cost per year. Average PCY sharply increases when $A_{optimal}$ < 100 sq ft. With the exception of one outlier, PCY stayed below $4000.00 for all air purifiers with $A_{optimal}$ > 100 sq. ft. The inverse relationship can be explained due to the normalization in the PCY metric. A single air purifier with low coverage will not be able to filter a 2500 foot space adequately; therefore, multiple purifier units are required. The extra cost generated by multiple units dramatically increases the overall cost, affecting the PCY.

**Figure 2**

*PCY as a Function of Optimal Coverage ($A_{optimal}$)*

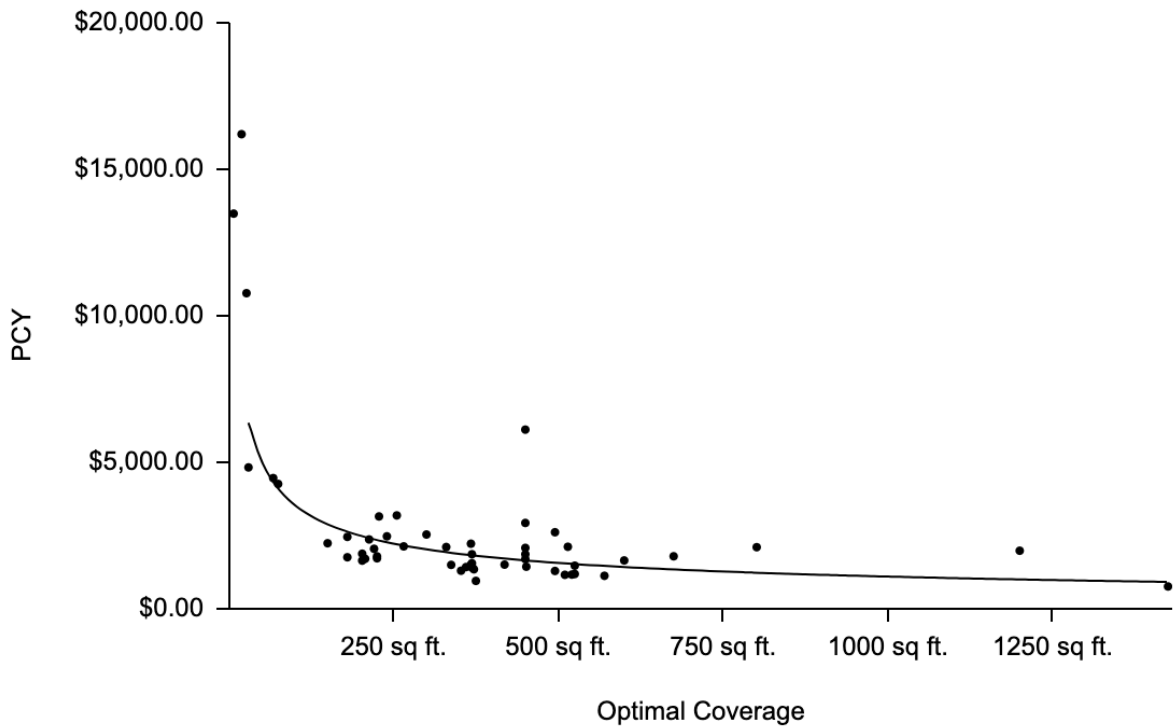

Optimal Coverage

PCY is not strongly correlated with initial cost or replacement filter cost. $C_{initial}$, $C_{maintenance}$, and PCY have coefficients of variation of 1.00, 0.74, and 1.06 respectively. A slight correlation between electricity cost and PCY was observed, but not enough data was gathered for any significance.

An interesting figure arises by relating $A_{optimal}$ to $C_{initial}$. Given the scope of this study, the trendline reveals that the two variables are nearly perfectly correlated when averaged.

**Figure 3**

*High correlation between purifier initial cost and optimal coverage*

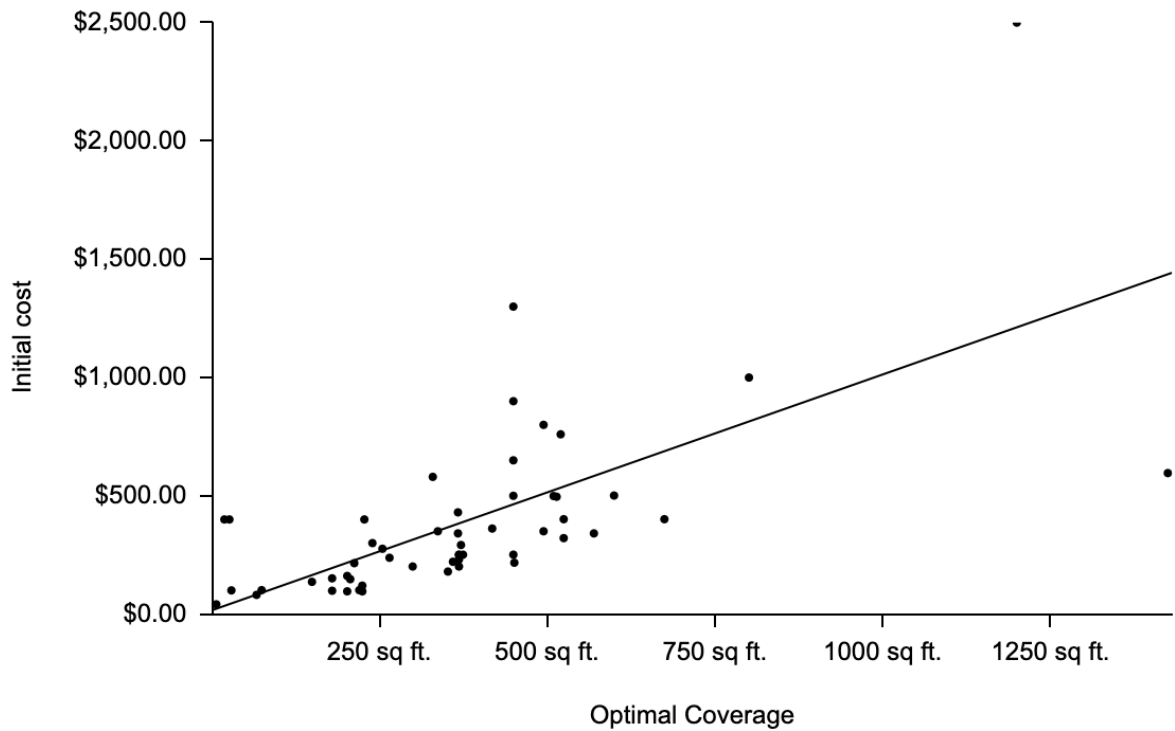

*Note.* The equation for the trendline is 0.998x + 13.4. The coefficient of determination $R^2=0.431$

Theoretically, the one-to-one nature of this relationship is quite convenient. Seeing as the $R^2$ is quite low, however, this observation can be used as an estimate at best.

### 4.3 Best Performing Purifier Units

Overall, the best performing models were the Medify MA-112, Blueair Pure 311 Auto, Blueair Pure 211+ Auto, Coway Airmega 250, and the Coway Airmega AP-1216L.

The PCY above assumes operating an air purifier continuously. A more optimized model involves running an air purifier solely on days with significant pollution, $PCY_{AQI}$. In the US, days with an AQI (air quality index) of over 100 are deemed "Orange Days", where sensitive groups such as asthmatics are at risk. Running the air purifier only on days with significant air pollution (i.e, an AQI of over 100) significantly decreases the usage time, and therefore decreases cost. AQI measurements vary with location. In our PCY calculation, $T_{operate}$ corresponds to the specific days. Based on data from the American Lung Association, 10 California counties with varying levels of particle pollution are analyzed leading to a median PCY of $506.09. The county with the highest number of bad air days, Kings County due to wildfires, averaged $694.01. In a moderately polluted county, Los Angeles, the average dropped to $347.01 with a median of $253.04.

### 4.4     Comparison with Medical Cost

Environmental decrease in PM 2.5 has been shown to lead to a decrease in medical conditions such as asthma attacks, especially for the elderly, infants, and people with weakened lungs [3][10][4]. While studies have been performed to assess the improvement of symptoms in vulnerable populations, few quantify the cost benefits of using an air purifier.  This section discusses asthmatic children, as a vulnerable group, and the yearly health cost they incur from pollution-related medical treatment in comparison to the PCY above

A study on Korean asthmatic children in 2020 indicated that through the use of air purifiers, the frequency of medication usage significantly decreased. Asthmatic children were monitored with and without daily air purifier usage. Participants were assessed daily for asthma severity, lung function, and most significantly, frequency of medication usage. In addition to noting a reduction in PM 2.5 levels, they concluded that air purifiers had a positive health effect on asthmatic children and a decrease in medication usage [5].

Similarly, a 2021 study focused on traffic-related air pollutants determined that HEPA filtration significantly helped asthmatic respiratory function. Asthmatic children living near major roadways aged between 10 to 16 had air purifiers placed in their bedrooms. Questionnaires and spirometry were utilized to test for respiratory health. They also showed that a decline in indoor traffic-related pollutants improved the quality of life for children with asthma [6].

According to a study in 2016, asthmatics incur $1990 per year in medical costs [4]. Comparing this cost to the PCY table shows that 47 out of 52 air purifiers run in highly polluted counties are below the cost threshold. Higher-performing units have a PCY well below the $1990 threshold. Given that the cost of air purification can be less than a quarter of the medical cost, air purifiers are a worthy investment.

## 5.  Conclusion

This study provides an objective evaluation of different air purifier units based on their overall cost. In a market where there is a myriad of models and products offered at a wide range in pricing, it is difficult to make purchasing decisions and to determine whether the cost is worthwhile. This article seeks to provide some insight into buying an air purifier and provides a metric for evaluation.

Several observations are made based on the metrics and data collected.

    (a) Within a given brand, a buyer should look for newer models. Newer models often work more efficiently, in addition to being more cost-efficient.

    (b) Replacement filters and electricity costs generate most of the cost of running air purifiers. The initial buying price only makes up 1/7 of the overall price on average.

    (c) To ensure adequate filtration of a certain room, the CADR should, at the very least, be at least ⅔ of the square footage of the room being filtered.

Finally, if the sampling in this study is representative of the air purifier market as a whole, investing in an air purifier is nearly always worth it. Given that the medical cost of a pollutant-sensitive population easily exceeds $1000 per year, the cost of purchasing and operating an air purifier is easily worthwhile.